\documentclass[onecolumn,pra]{revtex4}
\usepackage{amsmath,amsfonts,graphicx,color}

\newcommand{\ket}[1]{|#1\rangle}

\begin{document}

\title{Characterizing measurement-based quantum gates in quantum many-body systems using correlation functions} %
\author{Thomas Chung}%
\address{School of Physics, The University of Sydney,
  Sydney, New South Wales 2006, Australia}%
\author{Stephen D. Bartlett}%
\address{School of Physics, The University of Sydney,
  Sydney, New South Wales 2006, Australia}%
\author{Andrew C. Doherty}%
\address{School of Physical Sciences, The University of Queensland, St Lucia,
  Queensland 4072, Australia}%

\date{17 April 2009}

\begin{abstract}
  In measurement-based quantum computation (MBQC), local adaptive measurements are performed on the quantum state of a lattice of qubits.  Quantum gates are associated with a particular measurement sequence, and one way of viewing MBQC is that such a measurement sequence prepares a resource state suitable for `gate teleportation'.  We demonstrate how to quantify the performance of quantum gates in MBQC by using correlation functions on the pre-measurement resource state.
\end{abstract}

\maketitle

\section{Introduction}

A recent breakthrough in quantum computing has been the realization that quantum computation can proceed solely through single-qubit measurements on an appropriate quantum state~\cite{Rau01,Rau03}.  The canonical example of such a resource state is the cluster state~\cite{Rau01,Rau03}, which is a universal resource for MBQC on suitable lattices or graphs~\cite{vdN06}.  A handful of other universal resources for MBQC have recently been identified~\cite{Gro07a,Gro07b,vdN07,Bre08}, but it is still not known what properties of a quantum many-body system allow for MBQC to proceed.

In recent work~\cite{Doh08}, we have proposed that the ability to perform MBQC on a quantum many-body system may be identified using appropriate correlation functions as order parameters.  This claim stems from the observation that MBQC is a means of preparing resource states for gate teleportation~\cite{Got99,Chi05,Rau03}.  With a cluster state, it is possible by local measurements and feedforward alone to prepare such resource states allowing gate teleportation for a universal set of gates between essentially any set of qubits in the lattice.  The performance of MBQC can be determined by calculating the fidelity between the resource state that is actually prepared and the ideal resource state.   Here, we demonstrate how to express the resource state \emph{after} the measurements in terms of correlation functions of the original state \emph{prior} to the measurements.  These results provide an alternate perspective to Theorem 1 in~\cite{Rau03}, which shows that the gates in the cluster state MBQC scheme function because of certain correlations in the original cluster state.  In particular, our results apply to characterize gate performance in quantum states that are \emph{not} the cluster state; with such correlation functions, it is possible to directly quantify the suitability of a given quantum many-body state for performing such MBQC gates.

We note that our work has a close relation to the concept of localizable entanglement~\cite{Pop05}.  For the state of a quantum many-body system, the localizable entanglement between two arbitrary qubits is defined as the maximum amount of entanglement that can be created between these two qubits by performing local measurements on the remaining qubits.  If this entangled state is viewed as a resource for quantum teleportation, then the localizable entanglement serves to quantify the ability to perform the trivial or `identity' gate (i.e., teleportation) using local measurements.  We note that the localizable entanglement in some systems can be quantified by correlation functions~\cite{Pop05,Cam05}, using similar techniques as described here.  Our work generalizes these results by considering non-trivial quantum gates, which include multi-qubit quantum gates.

The paper is structured as follows.  In Sec.~\ref{sec:background}, we review some of the essential terminology and mathematical structure of cluster-state quantum computing.  Sec.~\ref{sec:general} presents the key general results of the paper, relating MBQC quantum gates to correlation functions.  The correlation functions for the identity gate are calculated explicitly in Sec.~\ref{sec:identity}, and non-trivial gates including the $\pi/2$-gate, Hadamard, and $Z$-rotation gates are presented in Sec.~\ref{sec:nontrivial}.  Two-dimensional gates are addressed in Sec.~\ref{sec:2D}, and a general method for concatenating gates in Sec.~\ref{sec:Concatenate}.  We finish with some brief conclusions in Sec.~\ref{sec:conclusions}.

\section{Pauli operators, stabilizers, and the cluster state}
\label{sec:background}

The Pauli matrices are labeled $X$, $Y$, $Z$ and are defined as
\begin{equation}
  X = \begin{pmatrix} 0& 1\\1& 0\end{pmatrix}\,, \qquad Y = \begin{pmatrix} 0& -i\\i& 0\end{pmatrix}\,, \qquad Z = \begin{pmatrix} 1& 0\\0& -1\end{pmatrix}\,.
\end{equation}
The group generated by the Pauli matrices under matrix multiplication is known as the Pauli group $G_1$. The Pauli group $G_n$ on $n$-qubits is defined as $G_n = G_1 \otimes G_1 \otimes \cdots G_1$.  The Clifford group on $n$ qubits is defined to be the group of unitary operators that map the Pauli group onto itself; i.e., $U$ is in the Clifford group if $UAU^{-1} \in G_n$ for all $A \in G_n$.

\subsection{Stabilizer Formalism}

The stabilizer formalism is a method of describing a state of a quantum system by specifying a set of eigenvalue relations instead of its components in some basis.  The eigenvalues of a complete set of commuting observables completely specifies a state.  We define the \emph{stabilizers} of a quantum state to be the set of operators for which the state is a $+1$ eigenstate.  Clearly, any two stabilizers must commute, and thus the set of stabilizers forms an Abelian group.  This group can be specified by its generators, and homomorphisms on the group can be specified completely by their effect on the generators. As a result it is sufficient to study the generators, a subset of the group, instead of the whole group.  For example, the state $\ket{00}$ can be described as the state which is stabilized by the operators $Z_1$ and $Z_2$; the stabilizer group of this state is generated by these two stabilizer operators.  The stabilizer formalism was first used to describe quantum error correction codes \cite{Got96}, but are widely applicable to a variety of other situations.

The standard stabilizer formalism is defined to only allow elements of the Pauli group $G_n$ to be used as stabilizers.  A set of Pauli stabilizers satisfies some key properties:
\begin{itemize}
    \item Elements of $G_n$ either commute or anti-commute.
    \item If $A, B, C \in G_n$ and $A$ anticommutes with both $B$ and $C$, then $A$ commutes with $BC$.
\end{itemize}

The power of the stabilizer formalism lies in its ability to compactly describe certain quantum states, as well as their evolution under Clifford group operations and Pauli measurements.  First, an arbitrary $n$-qubit state requires $2^n$ complex numbers to completely describe, by specifying the contribution from each of the $2^n$ basis vectors of the $n$-qubit system. For systems which are stabilized by the Pauli group, a set of $n$ stabilizers is sufficient to describe an $n$-qubit stabilizer state \cite{Got96}, i.e., $n$ stabilizers can form a complete set of commuting observables.  Hence if applicable, the stabilizer formalism offers a compact way of denoting states of a quantum system.  Second, the stabilizer formalism is very efficient in describing a state under the evolution of unitary operators belonging to the Clifford group, the group of operators which map the Pauli group back to itself under conjugation.  It is also efficient for describing the evolution of a state of a multi-qubit system under a projective measurement in the $X$, $Y$, or $Z$ basis. A simple prescription exists which tells us how to obtain the stabilizers of the post-evolution/measurement state from the pre-evolution/measurement state~\cite{nielsen2000a}.

\subsection{Cluster State}

The cluster state is a many-qubit entangled state~\cite{Rau03}.  Consider an arbitrary graph with a qubit at each vertex; the cluster state on this graph is characterized by the stabilizers
\begin{equation}
    K_a = X_a \otimes_{n\in {\rm nbhd}(a)}Z_n \,,
\label{cluster_state_stabilizer}
\end{equation}
for each qubit $a$ of the state, where ${\rm nbhd}(a)$ is the set of qubits adjacent to $a$ (via the graph structure).  In other words, the cluster state $|\Phi\rangle$ satisfies $K_a|\Phi\rangle = |\Phi\rangle$ for all $a$.

Two examples of cluster states on graphs will be considered here:  a one-dimensional line, for which the stabilizers take the form $K_i = Z_{i-1}X_i Z_{i+1}$, and a two-dimensional square lattice, for which the stabilizers take the form $K_i = X_{i,j} Z_{i,j-1}Z_{i,j+1}Z_{i-1,j}Z_{i+1,j}$.

\section{Expressing gates as correlation functions}
\label{sec:general}

Consider a one-dimensional lattice of qubits prepared in the state $\rho_0$.  Singling out two qubits, $a_{\rm in}$ and $a_{\rm out}$, we wish to consider a measurement sequence on the remaining qubits in the lattice that yields a two-qubit resource state on qubits $a_{\rm in}$ and $a_{\rm out}$ for gate teleportation.  Let $J$ label the measurement outcomes, and $P_J$ be the corresponding projector.  Following the measurements, a unitary $U_J$ conditional on $J$ is applied to qubit $a_{\rm out}$.  Averaged over all possible measurement outcomes, the resulting resource state is
\begin{equation}
  \rho= {\textstyle \sum_J} U_J P_J \rho_0 P_J U_J^\dagger\,.
\end{equation}
Equivalently, we can characterize this resource state using expectation values of bipartite Pauli operators $A \otimes B$ on qubits $a_{\rm in}$ and $a_{\rm out}$, as
\begin{equation}
  \langle A B\rangle =  {\textstyle \sum_J} {\rm Tr} [(AB)U_J P_J \rho_0 P_J U_J^\dagger ]
  =  {\textstyle \sum_J} {\rm Tr} [(AB_J)P_J \rho_0 P_J  ]\,,
\end{equation}
where $B_J = U_J^\dagger B U_J$.  The set of such correlation functions, for $A,B \in \{I,X,Y,Z\}$ spanning the set of Pauli operators, will completely specify the bi-partite resource state.

For each of the MBQC gates given in~\cite{Rau03}, we can make use of a remarkable relation:  there exists a string of operators $S$ acting on some set of the measured qubits which is \emph{independent} of the measurement outcomes, and an operator $B'$ on $a_{\rm out}$ that is also independent of $J$, such that
\begin{equation}
  \label{eq:CommutationIdentity}
  (AB_J)P_J = P_J A S B'\,.
\end{equation}
Now, using the projector properties $P_J^2 = P_J$ and $\sum_J P_J = I$ gives
\begin{equation}
  \label{eq:CorrFunction}
  \langle AB\rangle =  {\rm Tr} [(ASB') \rho_0]\,.
\end{equation}
Thus we can relate the resource state prepared \emph{after} the sequence of measurements to a correlation function of the original state $\rho_0$ \emph{prior} to measurements.  That is, the correlation functions characterize the \emph{post-}measurement resource state using expectation values of strings of operators on the \emph{pre-}measurement state.  This argument is trivially extendible to multi-qubit gates.

It is critical to this development that one can identify such a string of operators $S$.  In the examples presented, the correction operator is a Pauli operator, and essentially it is the simple algebraic properties of the Pauli operators which are responsible for the existence of $S$.  That the corrections are always Pauli operators makes the analysis of the Clifford gates especially simple, and we find that $S$ for such operators takes the form of a product of Pauli operators.  Even for our non-Clifford gate, we can still identify an appropriate operator (this time, a sum of product of Pauli operators).

\section{The Identity Gate}
\label{sec:identity}

We will consider performing an identity gate between two qubits in a line with an odd number of qubits between, qubit $k$ and qubit
$k+2l$.  To perform this gate one measures $X$ on each of the $2l-1$ qubits between these two, and measures $Z$ on
qubit $k-1$ and the qubit $k+2l+1$.  We label measurement outcomes by the eigenvalues of the measured operators, which for Pauli operators are either $+1$ or $-1$.  Specifically, we label the measurement outcome for the $X$ measurement on the qubit $k+j$ by $m_j=-1,1$, and the outcome for the $Z$ measurement on the qubit $k-1$ ($k+2j+1$) by $m_l$ ($m_r$).  We can then define two
parities
\begin{equation}
  p_Z = \Bigl[1-m_lm_r \bigl(\prod_{j=1}^{j=l-1} m_{2j} \bigr)\Bigr]/2 \,, \qquad
  p_X = \Bigl[1- \bigl(\prod_{j=1}^{j=l} m_{2j-1} \bigr)\Bigr]/2 \,,
\end{equation}
and write the correction unitary
\begin{equation}
  U_J= X_{k+2l}^{p_X} Z_{k+2l}^{p_Z}\,.
\end{equation}

We now seek to identify the commutation identity of Eq.~\eqref{eq:CommutationIdentity}. The unitaries only act on the $B$ space (qubit $k+2l$) so operators that act only on the $A$ space are unchanged on commuting through the projectors and do not depend on the measurement outcome.  For the $B$ operators $X_{k+2l}$ and $Z_{k+2l}$, we need to absorb the factors of $\pm 1$ that arise from the correction
unitary by adding factors of the measured observables. If we perform an $X$ measurement for example then $m P_m = P_m X$, where $P_m$ is the projector on the $m$ eigenstate.  We have
\begin{equation}
  X_{J} = U_J X_{k+2l} U_J^\dagger = m_l m_r \bigl(\prod_{j=1}^{j=l-1} m_{2j} \bigr) X_{k+2l}\,, \qquad
  Z_{J} = U_J Z_{k+2l} U_J^\dagger = \bigl(\prod_{j=1}^{j=l} m_{2j-1} \bigr) Z_{k+2l}\,,
\end{equation}
and therefore we have the relations
\begin{equation}
  X_J P_J = P_J Z_{k-1}\bigl(\prod_{j=1}^{j=l-1} X_{k+2j} \bigr) X_{k+2l} Z_{k+2l+1} \,, \qquad
  Z_J P_J = P_J \bigl(\prod_{j=1}^{j=l} X_{k+2j-1} \bigr) Z_{k+2l}\,.
\end{equation}

At this stage, we note that the correlation functions we would expect for an ideal identity gate are $\langle XX \rangle = \langle ZZ \rangle = - \langle YY \rangle = 1$.  Using Eq.~\eqref{eq:CorrFunction}, we find
\begin{align}
  \label{eq:XX}
  \langle XX\rangle &= {\rm Tr} \Bigl[ Z_{k-1} \bigl(\prod_{j=0}^{j=l} X_{k+2j}
  \bigr)  Z_{k+2l+1} \rho_0 \Bigr] = {\rm Tr} \Bigl[\bigl(\prod_{j=0}^{j=l} K_{k+2j}
  \bigr) \rho_0 \Bigr]\,, \\
  \label{eq:ZZ}
  \langle ZZ\rangle &= {\rm Tr} \Bigl[ Z_{k} \bigl(\prod_{j=1}^{j=l} X_{k+2j-1}
  \bigr)  Z_{k+2l+1}  \rho_0 \Bigr] = {\rm Tr} \Bigl[\bigl(\prod_{j=1}^{j=l} K_{k+2j-1}
  \bigr) \rho_0 \Bigr]\,.
\end{align}
The correlation function $\langle YY \rangle$ follows similarly.  As these correlations are equivalent to expectation values of cluster stabilizers, if $\rho_0$ is the cluster state these expectation values will both be unity.

Other expectation values, for example local expectation values $\langle XI\rangle$ or $\langle IX\rangle$, or those involving any other Pauli operator combination, can also be explicitly determined.  The set of all such correlation functions will completely characterize the resource state.

\section{Other single-qubit gates}
\label{sec:nontrivial}

We now turn our attention to other single-qubit gates.  Rather than directly consider correlation functions over arbitrary lengths, we restrict our attention to a fixed length (specifically, 3 intermediate qubits) between $a_{\rm in}$ and $a_{\rm out}$.  In Section~\ref{sec:Concatenate}, we will show how to concatenate such fixed-length gates together (possibly with the identity gate) to form gates of arbitrary lengths.  Specifically, we consider creating bipartite resource states between qubits $a_{\rm in}=1$ and $a_{\rm out}=5$ by measuring qubits 2, 3, and 4 along with $Z$ measurements on qubits 0 (to the left) and 6 (to the right).  See Fig.~\ref{fig:1Dgates}.

\begin{figure}
\begin{tabular}{|l|c|c|c|c|c|c|c|} \hline
    Identity & $Z$ & $\;\;\:$ & $X$ & $X$ & $X$ & $\;\;\:$ & $Z$ \\ \hline
    Hadamard & $Z$ & $\;\;\:$ & $Y$ & $Y$ & $Y$ & $\;\;\:$ & $Z$ \\ \hline
    $\pi$/2 & $Z$ & $\;\;\:$ & $X$ & $Y$ & $X$ & $\;\;\:$ & $Z$ \\ \hline
    $Z$-rotation & $Z$ & $\;\;\:$ & $X$ & $\pm\eta$ & $X$ & $\;\;\:$ & $Z$ \\ \hline
\end{tabular}
\caption{Measurement sequences to create the resource state for gate teleportation of various single-qubit gates.  The resource state qubits are $a_{\rm in}=1$ and $a_{\rm out}=5$, and qubits 0, 2, 3, 4, and 6 are measured as shown.}
 \label{fig:1Dgates}
\end{figure}

\subsection{The Hadamard gate}

Here we construct the correlation functions for the Hadamard gate $H=\frac{1}{\sqrt{2}}\begin{pmatrix} 1 & 1 \\ 1 & -1 \end{pmatrix}$. We perform the Hadamard gate, by measuring $Z$ on qubits 0 and 6 and measuring $Y$ on qubits 2, 3 and 4. This will give us correlation functions between qubits 1 and 5.  The ideal resource state for Hadamard gate teleportation satisfies $\langle XZ\rangle = \langle ZX\rangle = 1$.

Defining
\begin{equation}
  p_Z^H = \Bigl[1 - m_2 m_3 m_6 \Bigr]/2 \,, \qquad
  p_X^H = \Bigl[1 - m_0 m_3 m_4 \Bigr]/2 \,,
\end{equation}
the correction unitary for the Hadamard gate is then
\begin{equation}
  U_J = X_5^{p_X^H}Z_5^{p_Z^H}\,,
\end{equation}
We then have
\begin{equation}
  Z_JP_J = P_J Z_0 Y_3 Y_4 Z_5 \,, \qquad
  X_JP_J = P_J Y_2 Y_3 X_5 Z_6 \,,
\end{equation}
and therefore the correlation functions are given by
\begin{align}
  \langle XZ\rangle &= {\rm Tr}\bigl[Z_0 X_1 Y_3 Y_4 Z_5 \rho_0 \bigr] = {\rm Tr}\bigl[K_1 K_3 K_4 \rho_0 \bigr] \,, \nonumber \\
  \langle ZX\rangle &= {\rm Tr}\bigl[Z_1 Y_2 Y_3 X_5 Z_6 \rho_0 \bigr] = {\rm Tr}\bigl[K_2 K_3 K_5 \rho_0 \bigr]  \,.
\end{align}
We note that these correlation functions can be expressed as expectation values of products of cluster stabilizers; they will yield a value of one if $\rho_0$ is the cluster state.

\subsection{The $\pi/2$ gate}

The $\pi/2$ gate $S=\begin{pmatrix} 1 & 0 \\ 0 & i \end{pmatrix}$ is performed on a 1-D cluster state with qubit 1 as input and qubit 5 as output as in Fig.~\ref{fig:1Dgates}. The gate is implemented by measuring $Z$ on qubits 0 and 6, $Y$ on qubit 3 and $X$ on qubits 2 and 4. The ideal resource state for $\pi/2$ gate teleportation satisfies $\langle ZZ\rangle = \langle X(-Y)\rangle = 1$.

Defining
\begin{equation}
  p_X^{\pi/2} = \bigl[1 - m_2m_4\bigr]/2 \,, \qquad
  p_Z^{\pi/2} = \bigl[1 - m_0m_2m_3m_6\bigr]/2 \,,
\end{equation}
the correction unitary for the $\pi/2$ gate is then
\begin{equation}
  U_J = X_5^{p_X^{\pi/2}}Z_5^{p_Z^{\pi/2}}\,.
\end{equation}
The relevant correlation functions are then
\begin{align}
  \langle ZZ\rangle &= {\rm Tr}\bigl[Z_1 X_2 X_4 Z_5 \rho_0 \bigr] = {\rm Tr}\bigl[K_2 K_4 \rho_0 \bigr]\,, \nonumber \\
  \langle X(-Y)\rangle &= -{\rm Tr}\bigl[Z_0 X_1 Y_3 X_4 Y_5 Z_6 \rho_0 \bigr] = {\rm Tr}\bigl[K_1 K_3 K_4 K_5 \rho_0 \bigr]\,.
  \label{eq:pi2correlation}
\end{align}
Again, we note that these correlation functions can be expressed as expectation values of products of cluster stabilizers; they will yield a value of one if $\rho_0$ is the cluster state.

\subsection{A non-Clifford gate}

We now consider a non-Clifford gate -- a rotation $U_z(\theta)$ by angle $\theta$ about the $Z$ axis.  Again, we consider three intermediate qubits, as in Fig.~\ref{fig:1Dgates}.

The ideal resource state $\rho$ on $A$ and $B$ for gate teleportation of $U_z(\theta)$ satisfies $\langle Z Z \rangle = \langle X X_{-\theta} \rangle = 1$, where $X_{-\theta} = U_z(-\theta)XU_z(\theta) = \cos \theta X - \sin \theta Y$.  In MBQC, such a resource state is prepared on qubits 1 and 5 by measuring $Z$ on qubits 0 and 6, measuring $X$ on qubits 2 and 4, and measuring $X_\eta = U_z(\eta)X U_z(-\eta)$ on qubit 3, where $\eta = m_2 \theta$.  We note that the measurement basis on qubit 3 depends explicitly on the outcome of the measurement on qubit 2.  The correction unitary on qubit 3 is
\begin{equation}
  U_J = X_5^{p_X^{\theta}}Z_5^{p_Z^{\theta}}
\end{equation}
with
\begin{equation}
  p_X^{\theta} = \bigl[1 - m_2m_4\bigr]/2 \,, \qquad
  p_Z^{\theta} = \bigl[1 - m_0m_3m_6\bigr]/2 \,.
\end{equation}
For the $X$ measurements on qubits 2 and 4, we can use $m_2 P_J = P_J X_2$ and $m_4 P_J = P_J X_4$, and for qubits 0 and 6 we can use $m_0 P_J = P_J Z_0$ and $m_6 P_J = P_J Z_6$.  For the $X_\eta$ measurement yielding result $m_3$, the situation is slightly more complicated, because $\eta = m_2 \theta$.  However, it is straightforward to show that
\begin{equation}
  m_3 P_J = P_J (\cos\theta X_3 + \sin\theta X_2 Y_3)\,.
\end{equation}
Thus, we have
\begin{align}
  Z_J P_J &= P_J X_2 X_4 Z_5 \\
  X_J P_J &= P_J (\cos\theta Z_0 X_3 X_5 Z_6 + \sin\theta Z_0 X_2 Y_3 X_5 Z_6) \\
  Y_J P_J &= P_J (\cos\theta Z_0 X_2 X_3 X_4 Y_5 Z_6 + \sin\theta Z_0 Y_3 X_4 Y_5 Z_6) \,.
\end{align}
Note that the right hand side of the equations is of the desired form:  independent of the measurement results.

These results allow us to express the two-qubit expectation values for the post-measurement resource state in terms of correlation functions on the pre-measurement state.  We have
\begin{align}
  \langle Z Z \rangle &= {\rm Tr}[(Z_1 X_2 X_4 Z_5)\rho_0] = {\rm Tr}\bigl[K_2 K_4 \rho_0 \bigr] \\
  \langle X X_{-\theta}\rangle &= {\rm Tr}[(Z_0X_1X_3X_5Z_6(\cos^2\theta + \sin^2\theta Z_3X_4Z_5) \nonumber \\
  &\qquad\qquad + \cos\theta\sin\theta Z_0 X_1 X_2 Y_3 X_5 Z_6(1 -Z_3X_4Z_5))\rho_0] \nonumber \\
  &= {\rm Tr}[(K_1K_3K_5(\cos^2\theta + \sin^2\theta K_4) + \cos\theta\sin\theta (Z_0Y_1Z_2) K_2 K_3 (1-K_4)K_5)\rho_0] \,.
\end{align}

The term $Z_0Y_1Z_2$ is not a cluster state stabilizer, and has an expectation value of 0 for the cluster state. All other terms are stabilizers of the cluster state, and both correlation functions can be seen to have an expectation value 1 on the perfect cluster state.  In addition, this result agrees with Eq.~(\ref{eq:pi2correlation}) for $\theta=\pi/2$.

\section{Two-dimensional cluster states and the CSIGN gate}
\label{sec:2D}

The single-qubit gate sequences, and their corresponding correlation functions, can be straightforwardly generalized to the cluster state on a two-dimensional square lattice.  One-dimensional `strips' can be created in the square lattice by performing $Z$ measurements (and their corresponding Pauli corrections) to remove qubits from either side of the strip.

\begin{figure}
 \includegraphics[width=4in]{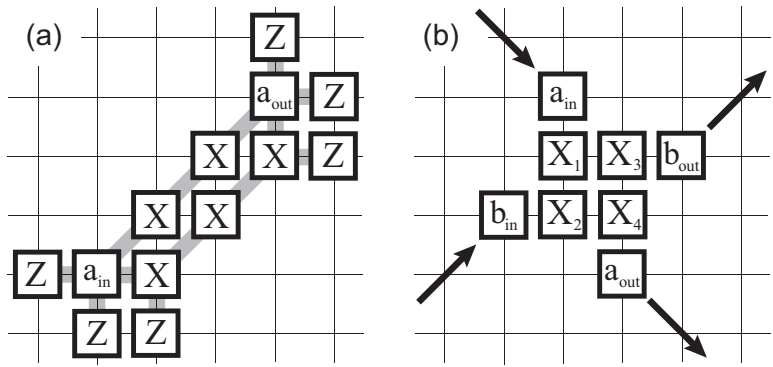}
\caption{(a) A measurement pattern on the cluster state that localizes entanglement between sites $a_{\rm in}$ and $a_{\rm out}$, where $X$ ($Z$) denotes a measurement in the $X$-basis ($Z$-basis).  The two strings of stabilizer products, centred on sites connected by the parallel diagonal lines, directly quantify the fidelities of single-qubit gates between $a_{\rm in}$ and $a_{\rm out}$ in MBQC.   (b)  The measurement sequence corresponding to the CSIGN gate between $a$ and $b$.  }
 \label{fig:2Dgates}
\end{figure}

However, for the purposes of defining simple correlation functions, it is easier to define single-qubit gate sequences along diagonal lines, as in Fig.~\ref{fig:2Dgates}(a).  Such diagonals eliminate the need for $Z$ measurements along the sides of the strip, and they are only required at the ends.  Consider an example where we label qubits such that $a_{\rm in}$ is at coordinate $(1,1)$ and $a_{\rm out}$ is at coordinate $(n,n)$.  For an ideal cluster state, the products of stabilizers $\prod_{i=1}^{i=n} K_{i,i}$ and $\prod_{i=1}^{i=n-1} K_{i+1,i-1}$ along and parallel to this diagonal are themselves stabilizers.  By measuring qubits $(i,i)$ for $1<i<n$ and qubits $(i+1,i-1)$ for $1\leq i \leq n-1$, we use the standard rules for updating stabilizers to give
\begin{align}
  \prod_{i=1}^{i=n} K_{i,i} &\rightarrow \Bigl[1-\textstyle{\prod_{i=2}^{i=n-1}}m_{i,i}\Bigr]X_{1,1}X_{n,n}Z_{0,1}Z_{1,0}Z_{n,n+1}Z_{n+1,n} \,,\\
  \prod_{i=1}^{i=n-1} K_{i+1,i-1} &\rightarrow \Bigl[1-\textstyle{\prod_{i=1}^{i=n-1}}m_{i+1,i-1}\Bigr] Z_{1,1}Z_{n,n} Z_{2,0} Z_{n+1,n-1} \,.
\end{align}
From this expression, we see that it is only necessary to measure qubits the six `end' qubits in the $Z$-basis in order to obtain the two-qubit resource state stabilized by $X_{1,1}X_{n,n}$ and $Z_{1,1}Z_{n,n}$; it is \emph{not} necessary to measure the qubits on either side of the diagonal strip.

With such diagonal strips, the correlation functions for the identity gate take the form
\begin{equation}
  \label{eq:ZX2D}
  \langle XX \rangle  = {\rm Tr}\Bigl[ \bigl( \prod_{i=1}^{j=n} K_{i,i} \bigr) \rho_0 \Bigr] \,, \qquad
  \langle ZZ \rangle  = {\rm Tr}\Bigl[ \bigl( \prod_{i=1}^{j=n-1} K_{i+1,i-1} \bigr) \rho_0 \Bigr] \,.
\end{equation}

We consider the simplest version of a two-qubit gate: the CSIGN gate (a Clifford gate) that also implements a cross-over of control and target qubits.  This gate is defined in terms of its action on Pauli operators as
\begin{equation}
  {\rm CSIGN}: \begin{cases} X_{a_{\rm in}} \otimes I_{b_{\rm in}} &\rightarrow X_{a_{\rm out}} \otimes Z_{b_{\rm out}} \\
    Z_{a_{\rm in}} \otimes I_{b_{\rm in}} &\rightarrow Z_{a_{\rm out}} \otimes I_{b_{\rm out}} \\
    I_{a_{\rm in}} \otimes X_{b_{\rm in}} &\rightarrow Z_{a_{\rm out}} \otimes X_{b_{\rm out}} \\
    I_{a_{\rm in}} \otimes Z_{b_{\rm in}} &\rightarrow I_{a_{\rm out}} \otimes Z_{b_{\rm out}} \end{cases}
\end{equation}
The measurement sequence is illustrated in Fig.~\ref{fig:2Dgates}.  The relevant correlation functions from the above `input-output' relations that will characterize the CSIGN gate are the expectations of four products of stabilizers:
\begin{align}
  \langle X_{a_{\rm in}} X_{a_{\rm out}} Z_{b_{\rm out}} \rangle &= {\rm Tr}\bigl[ K_{a_{\rm in}}K_3K_{a_{\rm out}} \rho_0\bigr]\,, \\
  \langle Z_{a_{\rm in}} X_{b_{\rm in}} X_{b_{\rm out}}  \rangle &= {\rm Tr}\bigl[ K_{b_{\rm in}}K_4 K_{b_{\rm out}} \rho_0\bigr]\,, \\
  \langle Z_{a_{\rm in}} Z_{a_{\rm out}} \rangle &= {\rm Tr}\bigl[ K_1 K_4 \rho_0\bigr]\,, \\
  \langle Z_{b_{\rm in}} Z_{b_{\rm out}}  \rangle &= {\rm Tr}\bigl[ K_2 K_3 \rho_0\bigr]\,,
\end{align}
can be appended with diagonal strings of stabilizers in the direction of the arrows (and terminated with $Z$ measurements as in Fig.~\ref{fig:2Dgates}(a)) to reach distant qubits.  With $X$ measurements on qubits 1-4, the resulting state provides the CSIGN transformation.

\section{Concatenating gates}
\label{sec:Concatenate}

It is straightforward to concatenate many Clifford gates together into a single gate, and to calculate the resulting correlation functions.  The essential idea is to equate the `output' qubit of the first gate with the `input' qubit of the second.  For the gates sequences defined here, recall that $Z$ measurements are performed at one qubit beyond each end of the gates. In this case where we combine two gates, the $Z$ measurement prior to the `input' end of the second gate and the `output' qubit of the first gate are not performed, and their corrections are to be left out.  Finally, an $X$ measurement is performed on this joining qubit.

As we have expressed the correlation functions of our Clifford gates in terms of cluster stabilizers, it is straightforward to determine the correlation functions describing a combined gate:  one simply takes the product of the corresponding stabilizer operators.

We note that it is still possible, though less straightforward, to concatenate non-Clifford gates.  The difficulty with non-Clifford gates is that they involve adaptive measurements, wherein the measurement basis can depend on the entire `past history' of the computation.  As a result, it is typically impossible to write the general form of the correlation functions for non-Clifford gates in a way that is independent of the specific choice of prior gates.  However, in any given situation, such correlation functions can be defined using the methods presented here.

\section{Conclusions}
\label{sec:conclusions}

We have presented a general method for expressing the performance of quantum gates in the cluster-state model of MBQC as correlation functions on the pre-measurement resource state.  With such correlation functions, viewed as order parameters, one can investigate the existence of a robust ordered phase in various models of quantum many-body systems that will allow for MBQC to occur.  One such model is considered in~\cite{Doh08}.

\section*{Acknowledgements}

This work is supported by the Australian Research Council.

\end{document}